\documentclass{llncs}
\usepackage{graphicx, amssymb}
\usepackage{llncsdoc} 

\begin{document}

\newcommand{\rrt}{{\tt RRT}}

\newcommand{\satz}{{\tt Satz}}
\newcommand{\zchaff}{{\tt zChaff}}
\newcommand{\walksat}{{\tt WalkSAT}}
\newcommand{\survey}{{\tt SP}}
\newcommand{\hgen}{{\tt hgen2}}
\newcommand{\e}{{\rm e}}
\newcommand{\os}{\overline{s}}
\newcommand{\remove}[1]{}

\title{From spin glasses to hard satisfiable formulas}
\author{Haixia Jia\inst{1}, Cris Moore\inst{1,2} and Bart Selman\inst{3} }
\institute{Computer Science Department, University of New Mexico, 
Albuquerque NM 87131 {\tt $\{$hjia,moore$\}$@cs.unm.edu} 
\and Department of Physics and Astronomy, University of New Mexico, 
Albuquerque NM 87131 
\and Computer Science Department, Cornell University, Ithaca NY
{\tt selman@cs.cornell.edu}}
\date{}
\maketitle

\begin{abstract}
We introduce a highly structured family of hard satisfiable 3-SAT formulas corresponding to an ordered spin-glass model from statistical physics.  This model has provably ``glassy'' behavior; that is, it has many local optima with large energy barriers between them, so that local search algorithms get stuck and have difficulty finding the true ground state, i.e., the unique satisfying assignment.  We test the hardness of our formulas with two Davis-Putnam solvers, \satz\ and \zchaff, the recently introduced Survey Propagation (\survey), and two local search algorithms, \walksat\ and
Record-to-Record Travel (\rrt). We compare our formulas to random 3-XOR-SAT formulas and to two other generators of hard satisfiable instances, the minimum disagreement parity formulas of Crawford et al., and Hirsch's \hgen. For the complete solvers the running time of our formulas grows exponentially in $\sqrt{n}$, and exceeds that of random 3-XOR-SAT formulas for small problem sizes. 
\survey\ is unable to solve our formulas with as few as 25 variables.  For \walksat, our formulas appear to be harder than any other known generator of satisfiable instances.  Finally, our formulas can be solved efficiently by \rrt\ but only if the parameter $d$ is tuned to the height of the barriers between local minima, and we use this parameter to measure the barrier heights in random 3-XOR-SAT formulas as well.

\end{abstract}

\section{Introduction}

3-SAT, the problem of deciding whether a given CNF formula with three literals per clause is satisfiable, is one of the canonical NP-complete problems. Although it is believed that it requires exponential time in the worst  case, many heuristic algorithms have been proposed and some of them seem to be quite efficient on average.  To test these algorithms, we need families of hard benchmark instances; in particular, to test incomplete solvers we need hard but {\em satisfiable} instances.  Several families of such instances have been proposed, including quasigroup completion~\cite{ssw,quasigroup1,quasigroup2} and random problems with one or more ``hidden'' satisfying assignments~\cite{Asahiro,VanGelder,AchJiaMoore}.  

In this paper we introduce a new family of hard satisfiable 3-SAT formulas, based on a model from statistical physics which is known to have ``glassy'' behavior.  Physically, this means that its energy function has exponentially many local minima, i.e., states in which any local change increases the energy, and which moreover are separated by energy barriers of increasing height.  In terms of SAT, the energy is the number of dissatisfied clauses and the global minimum, or ``ground state,'' is the unique satisfying assignment.  In other words, there are exponentially many truth assignments which satisfy all but a few clauses, which are separated from each other and from the satisfying assignment by assignments which dissatisfy many clauses.  Therefore, we expect local search algorithms like \walksat\ to get stuck in the local minima, and to have a difficult time finding the satisfying assignment.  

We start with a spin-glass model introduced by Newman and Moore~\cite{newmanmoore} and also studied by Garrahan and Newman~\cite{garrahannewman}.  It is like the Ising model, except that each interaction corresponds to the product of three spins rather than two; thus it corresponds to a family of 3-XOR-SAT formulas.  Random 3-XOR-SAT formulas, which correspond to a similar three-spin interaction on a random hypergraph and which are also known to be glassy, have been studied by Franz, M\'ezard, Ricci-Tersenghi, Weigt, and Zecchina~\cite{xorsatferro,xorsatsimplest,xorsatalternative},  Barthel et al.~\cite{xorsathiding}, and Cocco, Dubois, Mandler, and Monasson~\cite{xorsatexact}.  In contrast, the Newman-Moore model is defined on a simple periodic lattice, so it has no disorder in its topology.

We test our formulas against five leading SAT solvers: two complete solvers,  \zchaff\ and \satz, and three incomplete ones, \walksat,  \rrt\ and the recently introduced \survey.  We compare them with random 3-XOR-SAT formulas, and also with two other hard satisfiable generators, the minimum disagreement parity formulas of Crawford et al.~\cite{crawford} and Hirsch's \hgen\ \cite{hirsch}.  For Davis-Putnam solvers, our formulas are easier than random 3-XOR-SAT formulas of the same density in the limit of large size, although they are harder below a certain crossover at about $900$ variables.  For \survey, both our formulas and random 3-XOR-SAT formulas appear to be impossible to solve beyond very small sizes.  For \walksat, our formulas appear to be harder than any other known generator of satisfiable instances. We believe this is because our formulas' lattice structure gives them a very high ``configurational entropy,'' i.e., a very large number of local minima, in which local search algorithms like \walksat\ get stuck for long periods of time. 

The \rrt\ algorithm solves our formulas efficiently only if the parameter $d$ is set to the barrier height 
between local minima,  which for our formulas we know exactly to be $\log_2 L + 1$. 
Although the barrier height in random 3-XOR-SAT formulas seems to grow more quickly
with $n$ than in our glassy formulas, when $\sqrt{n} = L \le 13$ our formulas are harder for \rrt\ than random 3-XOR-SAT formulas of the same density, even when we use the value of $d$ optimized for each type of formula.  We propose using \rrt\ to measure barrier heights in other families of instances as well.

\section{The model and our formulas}

The Newman-Moore model~\cite{newmanmoore} consists of spins $\sigma_{i,j} = \pm 1$ on a triangular lattice.  Each spin interacts only with its nearest neighbors, and only in groups of three lying at the vertices of a downward-pointing triangle.  If we encode points in the triangular lattice as $(i,j)$, where the neighbors of each point are $(i \pm 1, j)$, $(i, j \pm 1)$, and $(i \pm 1, j \mp 1)$, the model's Hamiltonian (energy function) is 
\[ H = \frac{1}{2} \sum_{i,j}  \sigma_{i,j} \sigma_{i,j+1}\sigma_{i+1,j} \]  
Let us re-define our variables so that they take Boolean values, $s_{i,j} \in \{0,1\}$.  Then, up to a constant, the energy can be re-written
\[ H  = \sum_{i,j} (s_{i,j} + s_{i,j+1} + s_{i+1,j} ) \bmod 2 \]
In particular, we will focus on the case where the lattice is an $L \times L$ rhombus with cyclic boundary conditions; then
\[ H = \sum_{i,j=0}^{L-1} (s_{i,j} + s_{i,j+1 \bmod L} + s_{i+1 \bmod L,j} ) \bmod 2 \enspace . \]
Clearly we can think of this as a set of $L^2$ 3-XOR-SAT clauses of the form
\[ s_{i,j} \oplus s_{i,j+1 \bmod L} \oplus s_{i+1 \bmod L,j} = 0 \]
in which case $H$ is simply the number of dissatisfied clauses.  Each one of these can then be written as a conjuction of four 3-SAT clauses, 
\begin{eqnarray*}
(\os_{i,j} \vee s_{i,j+1 \bmod L} \vee s_{i+1 \bmod L,j}) 
& \wedge & (s_{i,j} \vee \os_{i,j+1 \bmod L} \vee s_{i+1 \bmod L,j}) \\
\wedge \;(s_{i,j} \vee s_{i,j+1 \bmod L} \vee \os_{i+1 \bmod L,j}) 
& \wedge & (\os_{i,j} \vee \os_{i,j+1 \bmod L} \vee \os_{i+1 \bmod L,j}) 
\end{eqnarray*}
producing a 3-SAT formula with $L^2$ variables and $4L^2$ clauses for a total of $12L^2$ literals.  
There is always at least one satisfying assignment, i.e., where $s_{i,j} = 0$ for all $i,j$.  However, using algebraic arguments~\cite{newmanmoore} one can show that this satisfying assignment is unique whenever $L$ has no factors of the form $2^m-1$, and in particular when $L$ is a power of $2$.  

To ``hide'' this assignment, we flip the variables randomly; that is, we choose a random assignment $A = (a_{i,j}) \in \{0,1\}^{L^2}$ and define a new formula in terms of the variables $x_{i,j} = s_{i,j} \oplus a_{i,j}$.  While some other schemes for hiding a random satisfying assignment in a 3-SAT formula create an ``attraction'' that allows simple algorithms to find it quickly, Barthel et al.~\cite{xorsathiding} pointed out that for XOR-SAT formulas these attractions cancel and make the hidden assignment quite difficult to find.  (Another approach pursued by Achlioptas, Jia, and Moore is to hide two complementary assignments in an NAESAT formula~\cite{AchJiaMoore}.)  Of course, XOR-SAT is solvable in polynomial time by Gaussian elimination, but Davis-Putnam and
local search algorithms can still take exponential time on random XOR-SAT formulas~\cite{xorsathiding,xorsatsimplest} 

\begin{figure}
\centerline{
\includegraphics[width=3.5in]{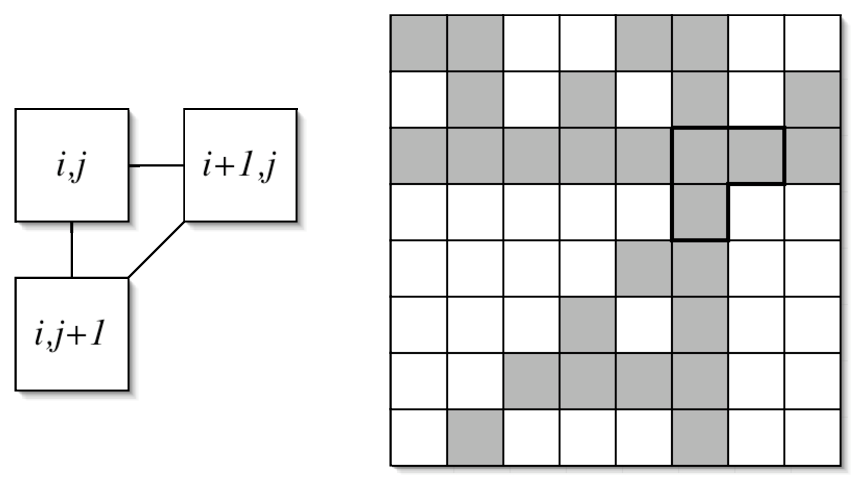}
}
\caption{A local minimum with a single defect.  Grey and white cells correspond to $s_{i,j} = 1$ and $0$ respectively; the XOR-SAT clause corresponding to the three cells outlined in black is dissatisfied, and all the others are satisfied.  The Hamming distance from the satisfying assignment is the number of grey cells, $L^{\log_2 3} = 27$ since $L=8$.}
\label{fig:defect}
\end{figure}

In general, XOR-SAT formulas have local minima because flipping any variable will dissatisfy all the currently satisfied clauses it appears in.  However, the lattice structure of the Newman-Moore model allows us to say much more.  In particular, if we call an unsatisfied XOR-clause a ``defect,'' then if $L$ is a power of $2$, there is exactly one state of the lattice for any choice of defect locations~\cite{newmanmoore}.  To see this, consider the state shown in Figure~\ref{fig:defect}.  Here there is a single defect (the three cells outlined in black) in which just one XOR-SAT clause (in fact, just one 3-SAT clause) is dissatisfied.  However, since satisfying the XOR-SAT clause at $i,j$ implies that
\[ s_{i,j+1} = s_{i,j} \oplus s_{i+1,j} \enspace , \]
the truth values below the defect are given by a mod-$2$ Pascal's triangle. If $L$ is a power of $2$ the $L$'th row of this Pascal's triangle consists of all $0$'s, so wrapping around the torus matches its first row except for the defect.  

This gives a truth assignment which satisfies all but one clause.  Moreover, this assignment has a large Hamming distance from the satisfying assignment; namely, the number of $1$'s in the Pascal's triangle, which is $H(L)=L^{\log_2 3}$ since it obeys the recurrence $H(2L) = 3 H(L)$.  It also has a large energy barrier separating it from the satisfying assignment: to fix the defect with local moves it is necessary to first introduce $\log_2 L$ additional defects~\cite{newmanmoore}.  

Now, by taking linear combinations (mod $2$) of single-defect assignments we can construct truth assignments with arbitrary sets of defects, and whenever these defects form an independent set on the triangular lattice, the corresponding state is a local energy minimum.  Thus the number of local minima equals the number of independent sets, which grows exponentially as $\kappa^{L^2}$ where $\kappa \approx 1.395$ is the {\em hard hexagon constant}~\cite{garrahannewman,Baxter}.  

To recap, when $L = 2^k$, there is a unique satisfying assignment.  The system is glassy in that there are many truth assignments which are far from the satisfying assignment, but which satisfy all but a small number of clauses.  Escaping these local minima requires us to first increase the number of unsatisfied clauses by roughly $\log_2 L$.  Newman and Moore~\cite{newmanmoore} studied the behavior of this model under simulated annealing, 
and found that the system is unable to find its ground state unless the cooling rate is exponentially slow; similarly, we expect the running time of local search algorithms like \walksat\ to be exponentially large.

Below, we compare our formulas to random satisfiable 3-XOR-SAT formulas, which were proposed in~\cite{xorsatsimplest} (and also in~\cite{xorsathiding} as the special case $p_0 = 1/4$).  These are formed with a random hidden assignment in the following way: given variables $x_1, \ldots, x_n$, select a random truth assignment $A \in \{0,1\}^n$.  Then, $m$ times, select a triple $x_i, x_j, x_k$ uniformly without replacement, and add the 3-XOR clause consistent with $A$, i.e. $x_i \oplus x_j \oplus x_k = a_i \oplus a_j \oplus a_k$.  To compare with our formulas, we set $n=m=L^2$ 
so the resulting 3-XOR-SAT formula has a density of one clause per variable.

\section{Experimental results}


\subsection{Davis-Putnam solvers: \zchaff\ and \satz}

\begin{figure}
\begin{center}
\includegraphics[width = 3in]{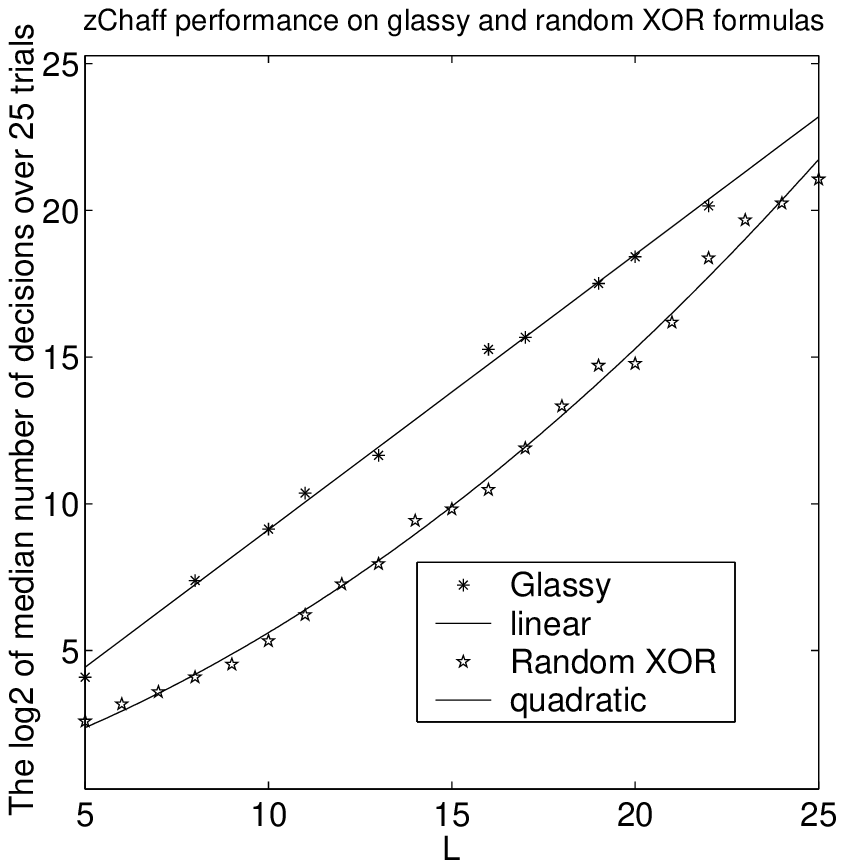}
\includegraphics[width =3in]{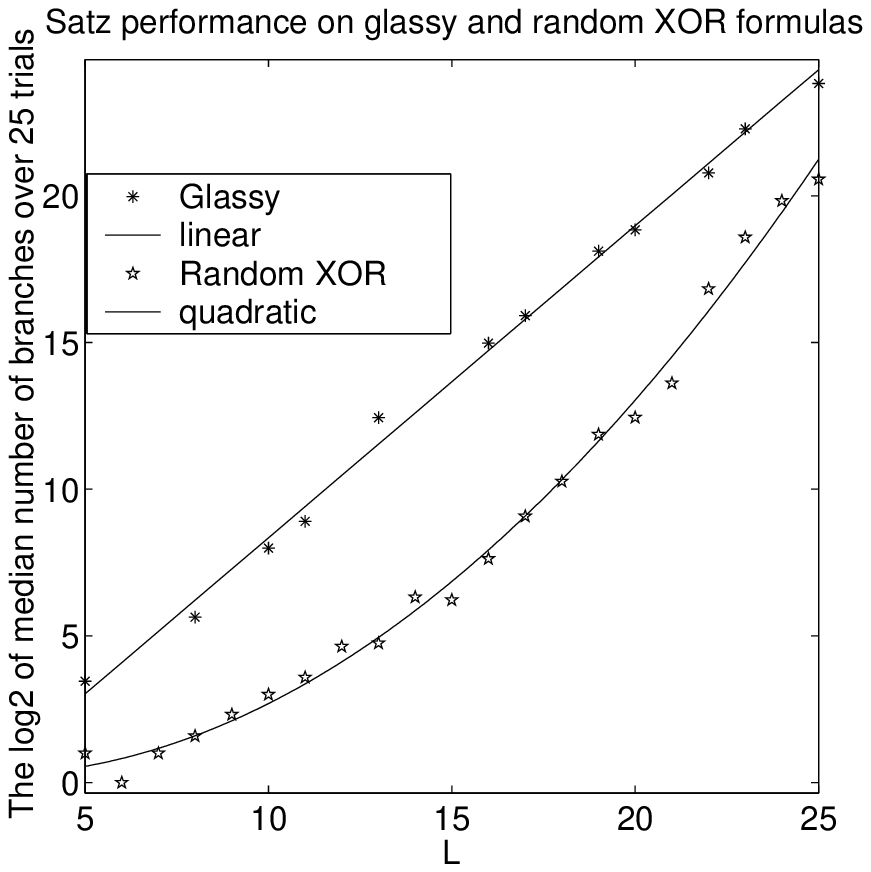}
\end{center}
\caption{The number of branches made by \zchaff\ and \satz\ on our formulas and on random 3-XOR-SAT formulas of the same size and density, as a function of the lattice size $L$.  The running time for random 3-XOR-SAT is exponential in $L^2 = n$, while for our formulas it is exponential in $L=\sqrt{n}$.  Nevertheless, for small values of $n$ our formulas are harder.  Each point is the median of $25$ trials; for our formulas, only values of $L$ for which the satisfying assignment is unique are shown.}
\label{fig1}
\end{figure}

We obtained \zchaff\ from the Princeton web site~\cite{zchaffsite} and \satz\ from the SATLIB web site \cite{satlib}.  Figure~\ref{fig1} shows a log-log plot of  the median number of decisions or branches that \zchaff\ and \satz\ took as a function of the lattice size $L$.  For both algorithms the slope for our glassy formulas is roughly $1$, indicating that the running time for \zchaff\ and \satz\ to solve our formulas grows as $2^L = 2^{\sqrt{n}}$.  The reason for this is that, due to a process similar to bootstrap percolation~\cite{bootstrap}, when a sufficient number of variables are set by the algorithm (for instance, the variables in a single row) the remainder of the variables in the lattice are determined by unit propagation.  For random 3-XOR-SAT formulas, the running time is exponential in $n=L^2$, but with a smaller constant, so that for $L \lesssim 30$ (i.e., $n \lesssim 900$) our formulas are harder than random 3-XOR-SAT formulas of the same size.

\subsection{\survey}

\survey\ is an incomplete solver recently introduced by M\'{e}zard and Zecchina~\cite{mz} based on a generalization of belief propagation called {\em survey propagation}.   For random 3-SAT formulas it is extremely successful; it can find a satisfiable assignment efficiently for random 3-SAT formulas up to size $n = 10^7$ near the satisfiability threshold $m/n \approx 4.25$ where random 3-SAT appears to be hardest.

We found that \survey\ cannot solve our formulas for $L \ge 5$, i.e., with $n=25$ variables.  The cavity biases continue to change, and never converge to a fixed point, so no variables are ever set by the decimation process.  There are several possible reasons for this.  One is the large number of local minima; another is that the symmetry in XOR clauses may produce conflicting messages; another is that our formulas have small loops which violate \survey's assumption that the formula is locally treelike and that neighbors are statistically independent.  
(Random 3-XOR-SAT formulas are also quite hard for \survey, although we found that \survey\ solved about $25\%$ of them with $n=m=25$.)



\subsection{Local algorithms: \walksat}

\begin{figure}
\begin{center}
\includegraphics[width = 3in]{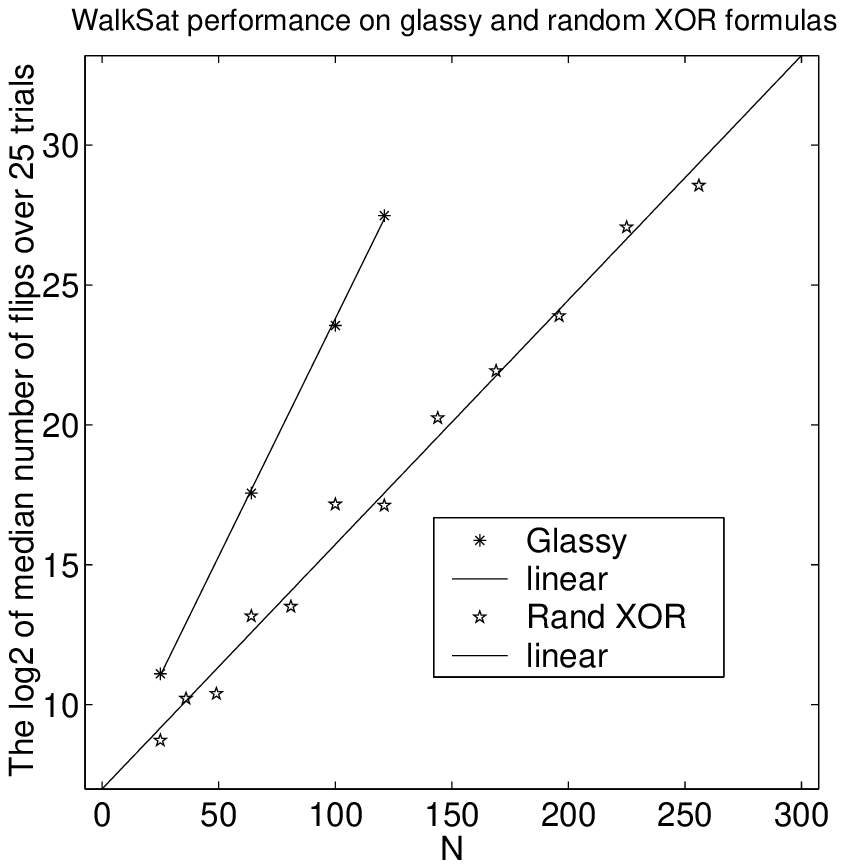}
\end{center}
\caption{The median number of flips made by \walksat\ on our formulas and random 3-XOR-SAT  formulas of the same size.  For our formulas, only values of $L$ for which the satisfying assignment is unique are shown.  Each point is the median of $25$ trials. }
\label{fig3}
\end{figure}

\walksat~\cite{walk} is an algorithm which combines a random walk search strategy with a greedy 
bias towards assignments with more satisfied clauses. \walksat\ has been shown to be highly 
effective on a range of problems, such as hard random k-SAT problems, graph coloring, and 
the circuit synthesis problem.  We performed trials of up to $10^9$ flips for each formula, without random restarts, where each step does a random or greedy flip with equal probability. Figure~\ref{fig3} shows a semi-log plot of the median number of flips as a function of $n=L^2$. We only choose four different values of $L$, namely $5$, $8$, $10$ and $11$, because \walksat\ was unable to solve the majority of formulas with larger values of $L$ (for which the satisfying assignment is unique) within $10^9$ flips. 

For both our formulas and random 3-XOR-SAT formulas, the median running time of \walksat\ grows exponentially in $n$.  However, the slope of the exponential is considerably larger for our formulas, making them much harder than the random ones.  We believe this is due to a larger number of local minima.  



\subsection{Local algorithms: \rrt}

\rrt~\cite{rrto,pekka} is a variant of \walksat\ which works as follows:
\begin{enumerate}
\item Start from a random truth assignment;
\item Randomly choose a variable from an unsatisfied clause; 
\item Flip it if this flip leads to a configuration that has at most $d$ more unsatisfied clauses than the best configuration found so far (the ``record''); don't flip otherwise; 
\item Repeat step 2 and 3 until it finds the satisfying truth assignment.
\end{enumerate}

Notice that $d$ is a predefined constant and will not be changed during the \rrt\ process.
In order to solve a formula, we have to find the ``right'' $d$ for \rrt. If we choose $d$ to be
too small, \rrt\ fails because it cannot escape the local minima;
and if we choose $d$ to be too big, 
it escapes the local minima but takes a long time to find the solution since
it is not greedy enough to move toward it.

\begin{figure}
\begin{center}
\includegraphics[width = 3in]{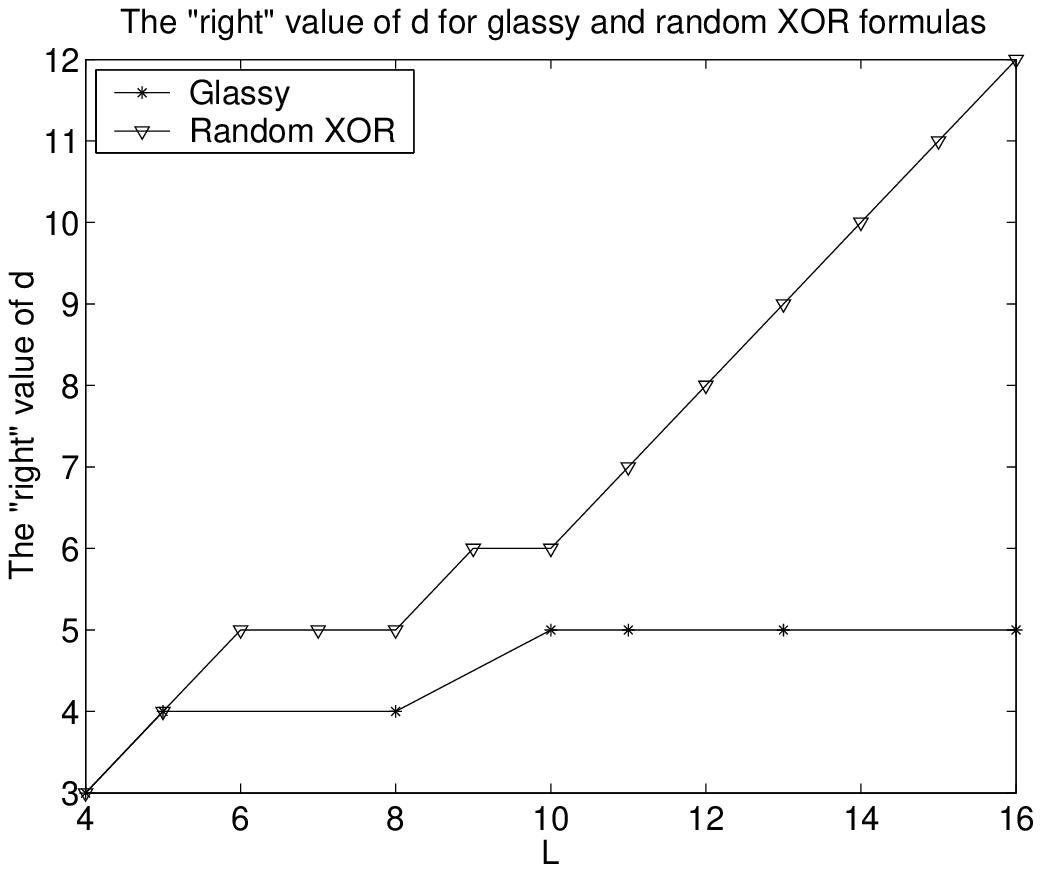}
\includegraphics[width =3in]{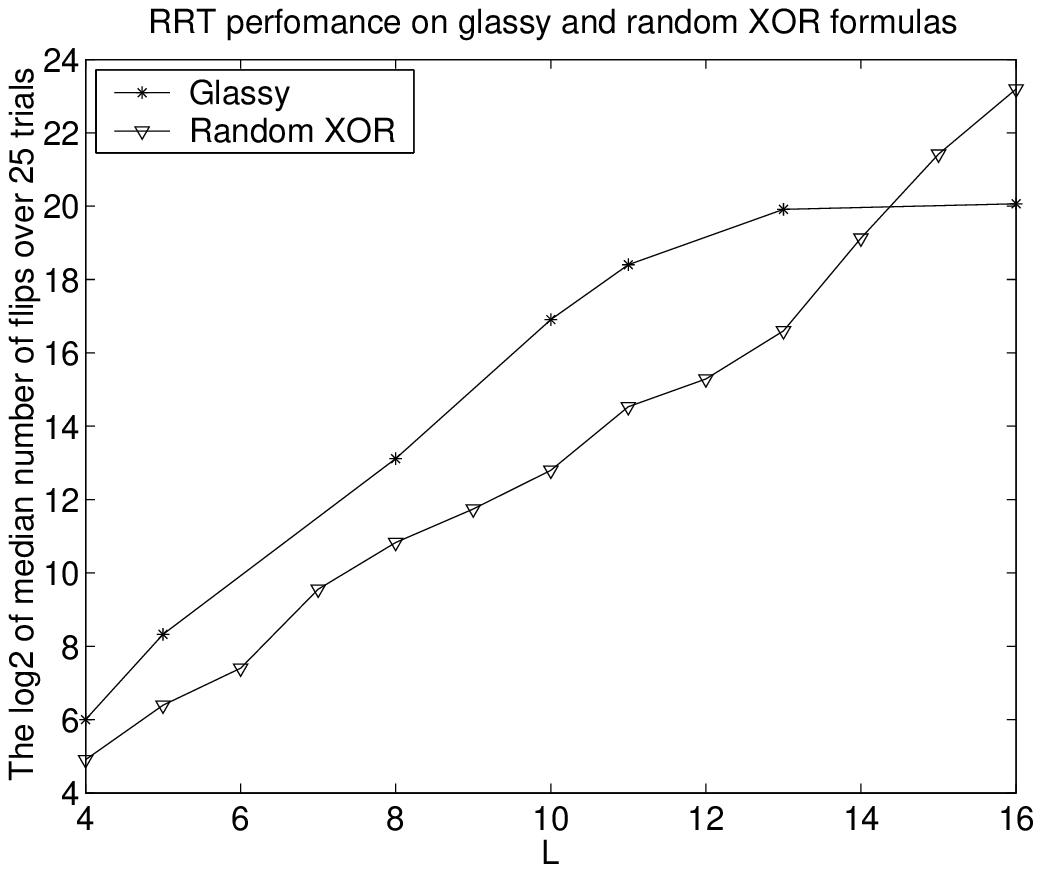}
\end{center}
\caption{The ``right'' value of $d$ and the running time for our formulas and random 3-XOR-SAT formulas of the same size.  Shown is the number of flips per variable.}
\label{rrt}
\end{figure}

We tested \rrt\ on our formulas with $L=4, 5, 8, 10, 11,13$ and $16$ and we performed
trials of up to $10^7$ flips for each formula. 
Newman and Moore~\cite{newmanmoore} showed that the largest barrier height 
is $\log_2 L+ 1$.  In fact, it turns out that \rrt\ solved our formulas efficiently only 
when $d = \log_2 L +1$ (see Figure~\ref{rrt}). With $L=16$
and $d=5$, \rrt\ solved our formulas in all of 50 trials with a median number of
flips $1.10 \times 10^6$; but when we set $d=4$ or $6$, \rrt\ can not solve any 
of the formulas with $L=16$ within
$10^7$ flips. \rrt\ may finds our glassy formula ``easy'' only if it knows the ``right'' value of $d$.

We also tested \rrt\ on random 3-XOR-SAT formulas with $n=m = L^2$ ranging 
from $16$ to $256$ so the resulting 3-XOR-SAT formula has same density as our glassy formulas.
Since we don't know the barrier height between local minima in these formulas, we tried 
\rrt\ with different values of $d$ to find the ``right'' $d$ for each value of $n$. 
As a rough measurement of the barrier heights, we measured the value $v$ for which 
\rrt\ solved more than half the formulas with $d=v$ but failed to solve half of them with $d=v-1$. 
We set the maximum running time to $10^7$ flips.

Figure~\ref{rrt} shows the ``right'' value of $d$ and the running time 
for each value of $n$. We see that
the barrier height in random 3-XOR-SAT formulas seems to grow more quickly
with $n$ than in our glassy formulas. 
However, when $\sqrt{n} = L \le 13$, our formulas are harder for \rrt\ than random 3-XOR-SAT formulas of the same density, even when we use the value of $d$ optimized for each type of formula.

We find it  interesting that \rrt\ can be used to measure the barrier heights between local minima, 
and we propose to do this for other families of formulas as well.

\subsection{Comparison with other hard SAT formulas}

To further demonstrate the hardness of our glassy formulas, we compare them to other two generators of hard instances: the parity formulas introduced by Crawford et al.~\cite{crawford} and the \hgen\ formulas introduced by E.A. Hirsch~\cite{hirsch}.  The parity formulas of~\cite{crawford} are translated from minimal disagreement parity problems and are considered very hard.  While \hgen\ does not generate parity formulas, we include it because it produced the winner of the SAT 2003 competition for the hardest satisfiable formula~\cite{satresult}.   

We compared our glassy formulas with $10$ formulas of Crawford et al., obtained from~\cite{crawford}, and with 25 \hgen\ formulas using the generator obtained from~\cite{hirsch}.   We ran \zchaff, \satz, \walksat\ and \survey; we did not test \rrt\ on these formulas.

For \walksat, we ran 25 trials of up to $10^9$ flips each, and labeled the formula ``not solved'' if none of these trials succeeded.  
Comparing our glassy formulas with those of Crawford et al., taking similar numbers of variables and clauses (e.g.\ comparing our $L=16$ formulas, which have $256$ variables and $3072$ clauses, with theirs with roughly $300$ variables and $4000$ clauses) we see from Table~1 that our formulas are significantly harder than theirs for \zchaff, \satz, and \walksat.  (\survey\ didn't solve any of these formulas, so it doesn't provide a basis for comparison.)  Compared to \hgen\ formulas with $195$ variables and $3096$ clauses, our formulas are not as hard for the complete solvers, but appear to be harder for \walksat, again perhaps due to their large number of local minima.

\begin{table}[ht]
\begin{center} {\footnotesize
\begin{tabular}{|c|c|c|c|c|c|}
\hline
Formulas & \multicolumn{1}{c|}{Literals } & \multicolumn{1}{c|}{Variables} &\multicolumn{1}{c|}{Dec. (\zchaff)} & \multicolumn{1}{c|}{Bran. (\satz)} &
 \multicolumn{1}{c|}{Flips (\walksat)} \\
\hline
par8-1-c.cnf & $732$ & 64 & 17 & 3  & 1494   \\
par8-2-c.cnf & $780$ & 68 & 9 & 1  & 2371   \\
par8-3-c.cnf & $864$ & 75 & 18 & 4  & 5638   \\
par8-4-c.cnf & $768$ & 67 & 7 & 1  & 2811   \\
par8-5-c.cnf & $864$ & 75 & 12 & 3  & 4828   \\
par16-1-c.cnf & $3670$ & 317 & 2073 & 1591  & $2.5\times10^8$   \\
par16-2-c.cnf & $4054$ & 349 & 11117 & 499  & $1.3\times10^8$   \\
par16-3-c.cnf & $3874$ & 334 & 7505 & 1489  & $1.0\times10^8$  \\
par16-4-c.cnf & $3754$ & 324 & 2181 & 4415  & $1.4\times10^8$  \\
par16-5-c.cnf & $3958$ & 341 & 2758 & 1296  & $4.1\times10^8$   \\ \hline
Glassy $8 \times 8$ & $768$ & 64 & 167 & 50 & 219455 \\
Glassy $16 \times 16$ & 3072 & 256 & 39293 & 32219 & not solved \\ \hline
Random XOR & $768$ & 64 & 23 & 3 & 9167 \\
Random XOR & $3072$ & 256 & 1427 & 198 & $3.9 \times 10^8$ \\
 \hline
\hgen & $3096$ & 295 &  not solved & 1478340  & 751723 \\
\hline
 \end{tabular} }
 \end{center}
 \caption{Comparison of our glassy lattice formulas with the parity formulas of Crawford et al., Hirsch's \hgen, and random 3-XOR-SAT formulas.}
 \label{turns}
\end{table}

\section{Conclusion}

We have introduced a new generator of hard satisfiable SAT formulas derived from a 
two-dimensional spin-glass model. We tested our formulas against five leading SAT solvers, and compared them with random 3-XOR-SAT formulas, the minimal disagreement parity formulas of Crawford et al., and Hirsch's \hgen\ generator.  For complete solvers, the running time of our formulas grows exponentially only in $L=\sqrt{n}$, but they are harder than random 3-XOR-SAT formulas when $n$ is small.  For \survey\ our formulas appear to be impossible for $n \ge 25$ variables.  
For \walksat\ our formulas appear to be harder than any other known generator of satisfiable instances. 
Finally, the \rrt\ algorithm solves our formulas only if $d$ is set to the barrier height between local minima, which we know exactly to be $\log_2 L + 1$.  We propose that \rrt\ can be used to measure the barrier heights between local minima in other families of instances, and we have done this for random 3-XOR-SAT formulas.

Since XOR-SAT is solvable in polynomial time, it would be interesting to have a provably glassy set of formulas which would be NP-complete to solve.  One approach would be a construction along the lines of~\cite{crawford}, where ``noise'' is introduced to the underlying parity problem so that it is no longer polynomial-time solvable.

Finally, we feel that the highly structured nature of our formulas, which makes it possible to prove the existence of exponentially many local optima with large barriers between them, suggests an interesting direction for future work.  For instance, are there families of formulas based on spin-glass models in three or more dimensions which would be even harder to solve?


{\bf Acknowledgments.}  We are grateful to Pekka Orponen for helpful discussions
on the \rrt\ algorithm and we are also grateful to the anonymous referees for proposing several directions for further work.  C.M. and H.J. are supported by NSF grant PHY-0200909
and H.J. is supported by NSF Graduate Research Fellowship.
C.M. is also grateful to Tracy Conrad for helpful discussions.

\end{document}